\documentclass[a4,aps,showpacs,amsmath,floatfix,twocolumn]{revtex4}
\usepackage{graphicx}
\usepackage{dcolumn}
\usepackage{color}
\usepackage{amsmath}
\usepackage{here}
\usepackage{bm}
\usepackage{amsmath}

\begin{document}
\title{p-wave superfluidity in mixtures of ultracold Fermi and spinor Bose gases}

\author{O.Y. Matsyshyn$^{1}$, A.I. Yakimenko$^{1}$, E.V. Gorbar$^{1,2}$, S.I. Vilchinskii$^{1}$, V.V. Cheianov$^{3}$}

\affiliation{$^1$  Department of Physics, Taras Shevchenko National University of Kyiv, 64/13, Volodymyrska Street, Kyiv 01601, Ukraine \\
$^2$ Bogolyubov Institute for Theoretical Physics, 14-b, Metrologichna, Kyiv 03680, Ukraine \\
$^3$ Instituut-Lorentz, Universiteit Leiden, P.O. Box 9506, 2300 RA Leiden, The Netherlands
}

\begin{abstract}
We reveal that the p-wave superfluid can be realized in  a mixture of fermionic and F=1 bosonic
gases. We derive a general set of the gap equations for gaps in the s-
and p-channels. It is found that the spin-spin bose-fermi interactions
favor the p-wave pairing and naturally suppress the pairing in the
s-channel. The gap equations for the polar phase of p-wave superfluid
fermions are numerically solved.  It is shown that a pure p-wave superfluid can be observed in a
well-controlled environment of atomic physics.
\end{abstract}

\pacs{05.30.Jp, 05.30.Fk, 03.75 Kk, 74.20.Rp}

\maketitle

\section{Introduction}
  Experimental realisation of a fermionic $p$-wave superfluid is one of the important challenges
 in ultracold atomic physics. If achieved in laboratory, it would mark an important milestone
 in the development of hardware for fault-tolerant quantum state manipulation
 \cite{RevModPhys-Das-Sarma}. Indeed, vortex excitations in a two-dimensional chiral version of
 such a condensate have been predicted \cite{ivanov2001non} to exhibit non-abelian braiding, which
 is the core underlying principle of topologically protected quantum computation \cite{Kitaev}.
 A few systems are already known today, which are believed to exhibit chiral $p$-wave pairing. These
 include the $\nu=5/2$ incompressible quantum Hall fluid \cite{moore1991nonabelions}, superfluid $^3$He at high pressure
 \cite{volovik2003universe}, and Strontium ruthenate \cite{kallin2012chiral}, and superconductor-topological insulator
 hybrid structures \cite{charpentier2017induced}. These systems however are not well
 suited for experimental realisation of quasiparticle braiding, which is mainly due to the limited
 local control of their parameters. Much hope is therefore devoted to the much better controlled
 ultracold atomic gases.

 First theoretical proposals for $p$-wave superfluidity in a ultracold fermionic gas
 were based around the idea of using Feshbach resonances in order to tune the $p$-wave
 collision channel into the regime of attraction \cite{Cooper}. Unfortunately, such resonant condensates
 were found to be extremely short lived due to three-body recombination and other decay mechanisms
 \cite{zhang2004p,gaebler2007p,levinsen2008stability,jona2008three}. Further theoretical work explored alternative mechanisms of pairing, which
 would be free of the disadvantages of a Feshbach resonance. A number of interesting proposals
 have emerged such as microwave dressed polar molecules \cite{cooper2009stable}, synthetic spin-orbit coupling
 \cite{zhang2008p,sato2009non,williams2012synthetic,julia2013engineering},
 the quantum Zeno effect \cite{han2009stabilization}, and driven dissipation \cite{bardyn2012majorana}. One important group of proposals
 rests on a traditional view that the nature of fermionic pairing depends on the nature of the agent,
 which mediates the attraction between fermions. In ultracold atomic systems it is natural to
 choose a Bose condensate as such a mediator. Possibility of a Cooper pairing in the $p$-wave channel in a
 fermion-boson mixture was originally discussed in \cite{Efremov}, albeit in an artificially
 spin-polarised system. More recent work explored pairing in a Fermi gas embedded in a spinless Bose condensate
 \cite{wang2005engineering,massignan2010creating}. In two recent papers \cite{wu2016topological,mathey2006competing} \
 it was proposed that a fermion-boson  system in mixed dimensions may exhibit an enhanced the $p$-wave pairing due to subtle correlation
 effects hidden in higher-order perturbation theory.

Here we explore a fermion-boson mixture in which the bosons form a polar spinor condensate.
We derive the effective action for the fermions and find that apart from the usual BCS-type
attraction excitations of the Bose condensate mediate a new type of interaction which
takes the form of spin-dependent X-Y exchange. We demonstrate that such an interaction between
the fermions may suppress pairing in the $s$-wave channel whilst favouring the p-wave pairing
channel. For a particular p-wave phase we derive and solve the complete set of Eliashberg
equations determining the gap parameter and we explore the dependence of the gap parameter
on the tunable parameters of the mixture.

\section{Model}

Let us describe {our} model of  a 3D {{spatially homogeneous}} mixture of electrically neutral  bosonic atoms with hyperspin $F=1$ and spin-$\frac12$ fermionic atoms. We
{assume} that an external magnetic filed is absent. The second quantized Hamiltonian of a mixture of dilute Fermi and spinor Bose
gases reads
\begin{equation}\label{HamiltTotal}
\hat H=\int\left[\mathcal{\hat H}_B(\textbf{r})+\mathcal{\hat H}_F(\textbf{r})+\mathcal{\hat H}_{BF}(\textbf{r})\right]d\textbf{r}.
\end{equation}
The Hamiltonian density $\mathcal{\hat H}_{B}(\textbf{r})$ describes {interacting} $F=1$ spinor bosonic atoms
\cite{PhysRevLett.81.742,2012PhR...520..253K,2013RvMP...85.1191S}
\begin{eqnarray}\label{HamiltB}\nonumber
\mathcal{\hat H}_{B}= -\frac{\hbar^2}{2M_B}\sum_m\hat{\Psi}_m^\dag\nabla^2
\hat{\Psi}_m
+ \frac12 c_0 : \hat n_B^2:+\frac12 c_2  : \hat{\textbf{F}} ^2 :,
\end{eqnarray}
where the field operators $\hat{\Psi}(\textbf{r})$ {obey} the bosonic comutation relations
$$
\left[\hat{\Psi}_m(\textbf{r}),\hat{\Psi}_{m'}^\dag(\textbf{r}')\right]=\delta_{m, m'}\delta (\textbf{r}-\textbf{r}')
$$
{and} $:\hat A:$ denotes normally ordered operator $\hat A$. {The $j$-th component of the vector spin density operator for bosons $\hat{\textbf{F}}$ equals}
$$
\hat{\textbf{F}}_j=\sum_{m, m'} \hat{\Psi}_m^\dag(\textbf{r}) \left\{ F_j\right\}_{m, m'}\hat{\Psi}_{m'}(\textbf{r}),
$$
where ${F}_j$ {are the $3\times 3$ matrices of the vector representation of the rotation group}. Further,
$$c_0=\frac{4\pi \hbar^2}{3 M_{B}}\left(a^{(0)}_{B}+ 2 a^{(2)}_{B}\right), \, c_2=\frac{4\pi \hbar^2}{3 M_{B}}\left(a^{(2)}_{B} - a^{(0)}_{B}\right)$$
are symmetric and spin-dependent interaction constants, respectively, $a^{(J)}_{B}$ is the scattering length in the state with spin $J$. We introduced also the number density
$\hat n_B(\textbf{r})=\sum_m \hat{\Psi}_m^\dag(\textbf{r}) \hat{\Psi}_m(\textbf{r})$, $m\in \{-1,0,+1\}$ of the bosonic  field. It is known that
the spin-1 BEC in the absence of an external magnetic field has two phases: ferromagnetic ($c_2<0$) and polar ($c_2>0$).
We consider {in this paper the polar spinor Bose--Einstein condensate (BEC) of $^{23}$Na atoms with $a_B^{(0)} = 50.0$,  $a_B^{(2)} = 55.0$ (see e.g. \cite{Pitaevskii2016}) in units of the Bohr radius. 

The {fermion} Hamiltonian density $\mathcal{\hat H}_{F}(\textbf{r})$ is given by
\begin{eqnarray}\label{HamiltF}\mathcal{\hat H}_F(\textbf{r})=\sum_\nu \hat{ f }_\nu^\dag\left(-\frac{\hbar^2}{2M_F}\nabla^2-\mu \right)
\hat{ f }_\nu
 + \frac12 g_F : \hat n_F^2 :,
\end{eqnarray}
where $\mu$ is the fermion chemical potential,  the {fermion field operators} $\hat{ f }$ obey the anticommutation {relations}
$$
\left\{\hat{ f }_\nu(\textbf{r}),\hat{ f }_{\nu'}^\dag(\textbf{r}')\right\}=\delta_{\nu, \nu'}\delta (\textbf{r}-\textbf{r}').
$$
Further, $\hat n_F(\textbf{r})=\sum_\nu \hat{ f }_\nu^\dag(\textbf{r}) \hat{ f }_\nu(\textbf{r})$, $\nu\in \{\uparrow,\downarrow\}$,
$g_F =  4\pi \hbar^2a_{F}/ M_{F}$ is the coupling constant of the fermion density-density interaction, and $a_{F}$ is the $s$-wave scattering length.

The interactions between bosons and fermions in the leading order in densities and spins  are described in Hamiltonian (\ref{HamiltTotal}) by the term
\begin{equation}\label{HBF}
\mathcal{\hat H}_{BF}(\textbf{r})=\frac{\alpha}{2}\, \hat n_B \hat n_F+\beta\, \hat{\textbf{S}}\cdot \hat{\textbf{F}},
\end{equation}
where $\alpha = {8\pi \hbar^2}\left(2a^{(1/2)}_{BF}+a^{(3/2)}_{BF}\right)/({3 M_{BF}}),$ and $ \beta = {8\pi \hbar^2}\left(a^{(1/2)}_{BF}-a^{(3/2)}_{BF}\right)/(3 M_{BF})$ are the density-density and spin-spin bose-fermi
interactions, respectively, $a^{(J)}_{BF}$ is the scattering length in the state with spin $J$, and $M_{BF}=2M_B M_F/(M_B+M_F)$ is the reduced
mass for a boson of mass $M_B$ and a fermion of mass $M_F$.

 The spin density operator for
fermions {has a similar form}
\begin{equation}
\hat{\textbf{S}}_j=\frac12\sum_{\nu, \nu'} \hat{ f }_\nu^\dag(\textbf{r}) \left\{ \sigma_j\right\}_{\nu, \nu'}\hat{ f }_{\nu'}(\textbf{r}),
\end{equation}
where $\hat\sigma_j$ are the Pauli matrices for spin-$\frac12$ fermions.

In what follows we neglect the back reaction of the fermion sector on the bose-condensate assuming that the bose-fermi interactions are weak.  


{In our study}, we  consider different values of bosonic density $n_B$ of $^{23}$Na at the fixed fermionic density $n_F = 10^{14}$ cm$^{-3}$ of  $^{40}$K atoms with the Fermi temperature 
 $T_F=E_F/k_B = 1248 $nK, where $E_F = {(3\pi^2 n_F)^{2/3}\hbar^2}/{(2 M_F)}$ is the Fermi energy.

\begin{figure}
\centering
\includegraphics[width=8.6cm]{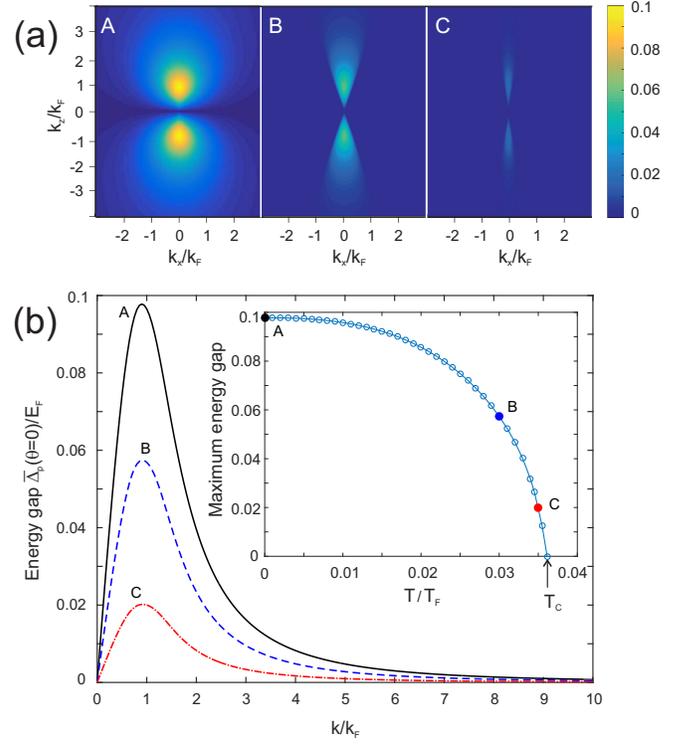}
\caption{Typical examples of the energy gap $\Delta = \Delta_p d_z$ in the p-wave polar phase found numerically at different temperatures
(A: $T=0$,
B: $T=0.03 T_F$, and C: $T=0.035 T_F$) for the interaction constants $\alpha =\beta = 0.3\,c_0$, $n_B=n_F= 10^{14}$ cm$^{-3}$: (a) The cross
section of the energy gap  $|\Delta(k_x, k_y=0, k_z)|$. (b) {The profile} of the energy gap along the $z$-axis. The
inset presents the maximum of the energy gap as a function of temperature. Note that the energy gap vanishes at the critical temperature
$T_c$.}
\label{fig1}
\end{figure}
\begin{figure*}
\centering
\includegraphics[width=\textwidth]{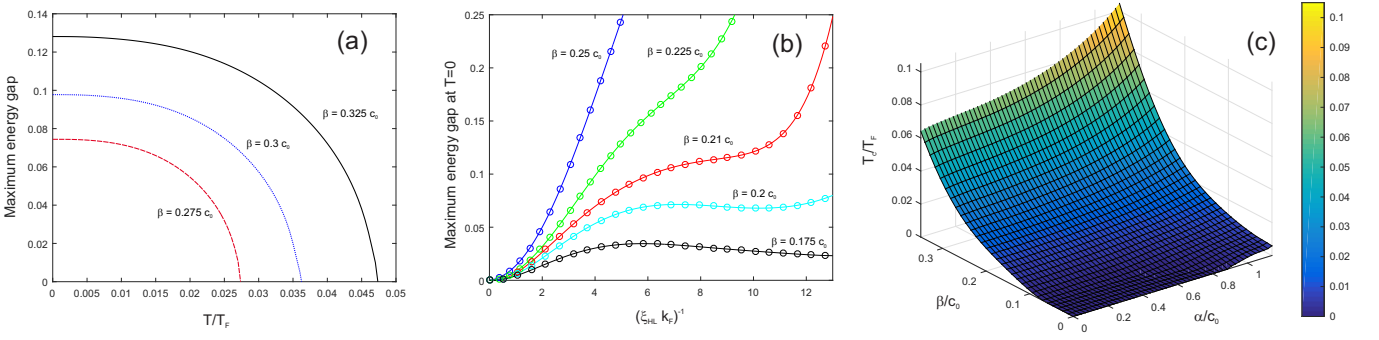}
\caption{(a) The maximal energy gap as a function of temperature for $\alpha = 0.3 c_0$ for different values of $\beta$ indicated near the curves. (b) The maximal energy gap at $T=0$ as a function of $(k_F \xi_\textrm{HL})^{-1}$ at fixed $\alpha = 0.3 c_0$ and $n_F=10^{14}$ cm $^{-3}$
is plotted for different values of $\beta$.  (c) The critical temperature versus the interaction
constants $\alpha$ and $\beta$ connected with $c_0$. The concentrations of bosons and fermions in (a) and (c) are fixed as follows:
$n_B=n_F= 10^{14}$ cm$^{-3}$.}
\label{fig2}
\end{figure*}

\section{Gap equations and solutions for polar superfluid phase}

Let us {show that the effective fermion-fermion interaction mediated by the interaction with $F=1$ bosons may give rise to
the $p$-wave} pairing between the fermions. Using Hamiltonian (\ref{HamiltTotal}), {we find that the fermion-boson interactions
lead to} the following effective fermion action:
\begin{equation}\label{eq62}
    S_{\text{eff}} = S_0 + S_{\textrm{int}},
\end{equation}
where

\begin{multline}
   S_0 = \int d\textbf{r}dt\Big[ \sum_{\nu} {f}^\dag_{\nu}\left(i  \frac{\partial}{\partial t} +\frac{\hbar^2}{2M_F}\nabla^2
   + \mu_\nu \right){f}_\nu - \\\frac{1}{2}g_F :  n_F^2(\textbf{r}) :\Big].
\end{multline}
is the free fermion action and the interaction term equals
\begin{multline}
   S_{\textrm{int}} = -\sum_{\textbf{k}\neq0}\frac{\epsilon_{\textbf{k}}n_B}{2}\int d\omega\Big\{\frac{\beta^2 }{E_{\textbf{k},1}^2
   - (\hbar\omega)^2} [\sigma_+(k)\sigma_-(-k)\\
    +\sigma_-(k)\sigma_+(-k)]
   +\frac{\alpha^2  }{2 (E_{\textbf{k},0}^2 - (\hbar\omega)^2)}n_F(k)n_F(-k)\Big\}.\\
\end{multline}
Here  $\epsilon_{\textbf{k}} = {\hbar^2\textbf{k}^2}/({2 M_B}$), $k = (\omega , \bf{k})$, $r = (t , \bf{r})$,
{$\Omega$ is system's volume},
\begin{equation}\label{sigmaPM}
   \sigma_\pm=\frac{1}{2\sqrt{\Omega}}\int e^{i {kr}}  f^\dag({r})({\sigma}_x\pm i {\sigma}_y)
   f ({r})d {r},\\
   \end{equation}
{and}
   \begin{equation}\label{BdG1}
     E_{\textbf{k},0} = \sqrt{\epsilon_{\textbf{k}} (\epsilon_{\textbf{k}} + 2c_0 n_B)},
   \end{equation}
   \begin{equation}\label{BdG2}
   E_{\textbf{k},\pm 1} = \sqrt{\epsilon_{\textbf{k}} (\epsilon_{\textbf{k}} + 2c_2 n_B)}
   \end{equation}
{are the Bogolyubov dispersion relations of the Bose gas excitations}.

Using the effective action \eqref{eq62}, we find the following Dyson's equations in the Matsubara formalism {for the fermion Green
function $G(\textbf{k},\omega_n)$ and the anomalous Gor`kov function $F(\textbf{k},\omega_n)$}:
\begin{equation}\label{Geq3}
    (i\hbar\omega_n-\xi_{\textbf{k}})G(\textbf{k},\omega_n)+\Delta(\textbf{k},\omega_n)F^\dag(\textbf{k},\omega_n) = \hbar,
\end{equation}
\begin{equation}\label{F+eq3}
    (i\hbar\omega_n+\xi_{\textbf{k}})F^\dag (\textbf{k},\omega_n)+\Delta^\dag(\textbf{k},\omega_n)G(\textbf{k},\omega_n)=0,
\end{equation}
where $\xi_k = {\hbar^2\textbf{k}^2}/({2M_F})-\mu$. Here
$\omega_n = {k_B T}(2n+1)\pi/{\hbar}$ is the Matsubara fermion frequency. In addition, the gap function $\Delta(\mathbf{k},\omega_n)$
neglecting the retardation effects, is expressed through the Gor`kov function as follows:
\begin{multline}\label{GAP}
\Delta(\textbf{k})
    =-\frac{k_B T}{\hbar}\sum_{n^\prime,\textbf{k}'\neq 0}\Big[ \frac{g_F}{\Omega}F(\textbf{k}-\textbf{k}',\omega_{n^\prime}) - \\
    \tilde{g}_n (\textbf{k}',\omega_{n^\prime})  F(\textbf{k}-\textbf{k}',\omega_n - \omega_{n^\prime}) \\
    -\tilde{g}_s (\textbf{k}',\omega_{n^\prime}) \sum_{s = +,-}\hat{\sigma}_s F(\textbf{k}-\textbf{k}',\omega_n - \omega_{n^\prime})\hat{\sigma}_s
    \Big],
\end{multline}
where
\begin{align*}
    \tilde{g}_n(\textbf{k},\omega_{n^\prime}) =  \frac{c_n}{ E_{\textbf{k},0}^2 + (\hbar\omega_{n^\prime})^2} , &\qquad
    c_n = \frac{\alpha^2 n_B \epsilon_{\textbf{k}}}{2\Omega},\\
    \tilde{g}_s (\textbf{k},\omega_{n^\prime})=  \frac{c_s}{ E_{\textbf{k},1}^2 + (\hbar\omega_{n^\prime})^2} ,&\qquad
    c_s = \frac{\beta^2 n_B \epsilon_{\textbf{k}}}{\Omega}.
\end{align*}

An anzatz for the energy gap, which accounts for both $s$- and $p$- wave channels, is given by
\begin{equation}\label{gapsi}
    \Delta_{\alpha \beta} (\textbf{k}) = \Delta_p (i\bm{\sigma} \sigma_y)_{\alpha \beta} \textbf{d}(\textbf{k})
    + \Delta_s (i\sigma_y)_{\alpha\beta},
\end{equation}
where $\Delta_p$ and $\Delta_s$ are real functions of $|\textbf{k}|$. As is well known, the order parameter of the p-wave phase is
proportional to vector $\textbf{d}$ whose form and symmetries determinate the superfluid phase.

It is straightforward now to obtain the set of gap equations for $\Delta_s$ and $\Delta_p$ using the standard procedure (see Appendix). There
is, {in principle, many solutions to the gap equations. We follow in our analysis the well known classification of
the $^3$He phases \cite{Min} defined by the form of vector $\textbf{d}$}. In the present work, we restrict our consideration to the polar phase
with $\textbf{d}\sim(0,0,k_z)$, which corresponds to one of the simplest solutions of the gap equations.
It turns out that {the} $s$-wave superfluidity can be suppressed by spin-induced interactions (see Appendix), thus, we assume for
simplicity that $\Delta_s=0$ and $\Delta_p\ne 0$. Then the gap equation in the polar phase reads
\begin{multline}\label{dz}
    {\Delta} = \\ = \sum_{\textbf{k}^\prime\neq0} \Big\{c_n\left(\frac{W_0}{2E}\tanh{\frac{E}{2k_BT}}
    +\frac{Z_0}{2E_{\textbf{k},0}}\tanh{\frac{E_{\textbf{k},0}}{2k_BT}}\right)+\\
    +c_s\left(\frac{W_1}{2E}\tanh{\frac{E}{2k_BT}}+\frac{Z_1}{2E_{\textbf{k},1}}\tanh{\frac{E_{\textbf{k},1}}{2k_BT}}\right)\Big\}\times \\
    \times {\Delta}(\textbf{k}-\textbf{k}^\prime),
\end{multline}
where ${\Delta}(|\textbf{k}|)=|d_z|\Delta_p(|\textbf{k}|)$,
\begin{gather*}
    W_i = \frac{E_{\textbf{k},i}^2-E^2+\hbar^2\omega_0^2}{(E_{\textbf{k},i}^2-E^2)^2+2\hbar^2\omega_0^2(E_{\textbf{k},i}^2+E^2+\frac{\hbar^2\omega_0^2}{2})},\\
    Z_i = \frac{E^2-E_{\textbf{k},i}^2+\hbar^2\omega_0^2}{(E_{\textbf{k},i}^2-E^2)^2+2\hbar^2\omega_0^2(E_{\textbf{k},i}^2+E^2+\frac{\hbar^2\omega_0^2}{2})},
\end{gather*}
with $\omega_0 = {k_B T \pi}/{\hbar}$,  and  $E = \sqrt{\xi_{\bf k} + \Delta^2}$.

Let us discuss the numerical solutions of Eq. (\ref{dz}) obtained by {using a} stabilized iterative procedure \cite{Petviashvili86}.
Since the function $\Delta$ has an axial symmetry, it suffices to consider a cross section of its energy gap at $k_y=0$.
Figure \ref{fig1} presents typical examples of {this cross section} at fixed temperature and
interaction constants $\alpha$ and $\beta$. It is seen from Fig. \ref{fig1} that the maximum of the energy gap is localized near the points
$(0,0,\pm k_F)$, and the maximal energy gap monotonically decreases as temperature increases [see the inset in Fig. \ref{fig1} (b) and
Fig. \ref{fig2} (a)]. The energy gap in the $p$-phase vanishes at some critical temperature $T_c$. We investigated the dependence of the
critical temperature $T_c$ {on the ratios $\alpha/c_0$ and $\beta/c_0$ of  boson-fermion} interactions to the strength of the
boson-boson interaction $c_0$ [see Fig. \ref{fig2} (c)].
According to Fig. \ref{fig2} (c), the critical temperature $T_c$ can be significantly increased through {an} interplay of the
density-density and spin-spin interactions between fermions and spinor BEC.

The energy gap is a function of interaction constants $\alpha$, $\beta$, $c_0$,  $c_2$ and densities of the Bose- and Fermi-gases.
Figure \ref{fig2} (b) illustrates how the energy gap is affected by variation of the Bose-gas density.
The non-trivial dependence of the maximal gap $\Delta_p$ on $(k_F\xi_\textrm{HL})^{-1}$, where $ \xi_\textrm{HL}=1/\sqrt{
({8\pi} n_B (a_B^{(0)}+2 a_B^{(2)})/3}
$, at $T=0$ comes from the Bogolyubov dispersion
relations (\ref{BdG1}) and (\ref{BdG2}) of the Bose gas. Such a dependence should be taken into account in the experimental observation of the $p$-wave superfluid state.

\section{Conclusions}

We showed that a mixture of dilute ultracold $F=1$ spinor atomic BEC and fermionic atoms provides a promising platform for the
realization of the $p$-wave superfluid state in three dimensions. Such a mixture makes possible to avoid the problem of three-body
inelastic collisional loss in a single-component ultracold Fermi gas, where the Feshbach resonance is used  to enhance the $p$-wave
cross section to values larger than the $s$-wave cross section. We derived a general set of the gap equations for the energy gaps
in the $s$- and $p$-channels. It is found that the spin-induced interactions in a mixture of $F=1$ spinor atomic BEC and fermionic atoms
naturally suppress the $s$-wave pairing. At the same time the $p$-wave pairing is driven not only by density-induced interaction, but also by spin-induced interactions and it emerges even for repulsive interaction between fermions. 

We studied the simplest case of the $p$-wave triplet pairing defined by the polar phase with $\textbf{d}=(0,0,k_z)$. For the case of
dilute Fermi $^{40}$K and Bose $^{23}$Na atomic gases, we found numerically
the solution to the gap equations of the polar phase with the vanishing $s$-wave order parameter $\Delta_s=0$. While the dependence of
the maximal value of the gap on temperature is a monotonically decreasing function, its dependence on the 
density of bosons $n_B$ is non-monotonous for sufficiently small values of the spin-spin bose-fermi interaction $\beta$. Such a dependence comes
from the corresponding non-trivial dependence of the Bogolyubov dispersion relations on $n_B$.

We hope that the results presented in this paper will stimulate experiments to observe the $p$-wave superfluid in a {mixture of
ultracold Fermi and spinor Bose gases}. Moreover, the study of the $p$-wave superfluidity in a well-controlled environment of atomic physics
could help to elucidate the properties of {topologically non-trivial superfluids}.

\section*{ACKNOWLEDGMENTS}
O.M. and A.Y. acknowledge {the support of the} Project 1/30-2015 'Dynamics
and topological structures in Bose-Einstein condensates of
ultracold gases' of the KNU Branch Target Training at the
NAS of Ukraine. A.Y. acknowledges the hospitality of the Leiden University, where this work was commenced.
{The work of E.V.G. was partially supported by the Program of Fundamental Research of the Physics and Astronomy Division of the
National Academy of Sciences of Ukraine}.

\section*{Appendix.  General gap equations}
Using anzatz (\ref{gapsi}) we rewrite Eqs. \eqref{Geq3} and \eqref{F+eq3} as
\begin{multline}\label{solGPS}
    G_{\alpha\beta}(\textbf{k},\omega_n) = - \hbar(i\hbar\omega_n +\xi_{\textbf{k}})\times \\ \times\frac{(\hbar^2\omega_n^2+\xi_{\textbf{k}}^2
    +\Delta^2)\delta_{\alpha\beta} - \Delta^2\textbf{M}\bm{\sigma}_{\alpha\beta}}{(\hbar^2\omega_n^2 + E_+^2)(\hbar^2\omega_n^2 + E_-^2)},
\end{multline}
\begin{multline}\label{solFPS}
    F_{\alpha\beta}(\textbf{k},\omega_n) = \hbar\Delta_p (\bm{\sigma}i\sigma_y)_{\alpha\beta}\times\\ \times\frac{(\hbar^2\omega_n^2
    +\xi_{\textbf{k}}^2+\Delta^2)\textbf{d} - i\Delta^2[\textbf{M}\times\textbf{d}]}{(\hbar^2\omega_n^2 + E_+^2)(\hbar^2\omega_n^2 + E_-^2)} +\\
    +\hbar\Delta_s\frac{(\hbar^2\omega_n^2+\xi_{\textbf{k}}^2+\Delta^2-2\Delta_p^2(\textrm{Re}[\textbf{d}]\cdot\textbf{d}))i\sigma_{y,\alpha\beta} }
    {(\hbar^2\omega_n^2 + E_+^2)(\hbar^2\omega_n^2 + E_-^2)}+\\+\hbar\Delta_s\frac{ \Delta^2(\bm{\sigma}i\sigma_y)_{\alpha\beta}\textbf{M}}
    {(\hbar^2\omega_n^2 + E_+^2)(\hbar^2\omega_n^2 + E_-^2)},
\end{multline}
{where}
\begin{gather*}
    \Delta^2 = \Delta_p^2 |\textbf{d}|^2 +\Delta_s^2, \\
    \textbf{m}(\textbf{k}) = i [\textbf{d}(\textbf{k})\times \textbf{d}^*(\textbf{k})],\\
    \textbf{M} = \frac{\Delta_p^2\textbf{m} + 2 \Delta_p\Delta_s \textrm{Re}[\textbf{d}]}{\Delta^2},\\
    E_{\pm} =\sqrt{\xi_{\textbf{k}}^2 + \Delta^2 ( 1 \pm | M |) }.
\end{gather*}

{The general gap equations read}
\begin{multline}\label{solx}
    d_x(\textbf{k})\Delta_p(\textbf{k},\omega_k) = k_B T \times \\ \times\sum_{n,\textbf{k}'\neq0} \tilde{g}_n(\textbf{k}',\omega_k
    -\omega_n)\Delta_p(\textbf{k}-\textbf{k}',\omega_n)\upsilon_x(\textbf{k}-\textbf{k}',\omega_n),
\end{multline}
\begin{multline}\label{soly}
    d_y(\textbf{k})\Delta_p(\textbf{k},\omega_k) = k_B T \times \\ \times\sum_{n,\textbf{k}'\neq0} \tilde{g}_n(\textbf{k}',\omega_k
    -\omega_n)\Delta_p(\textbf{k}-\textbf{k}',\omega_n)\upsilon_y(\textbf{k}-\textbf{k}',\omega_n),
\end{multline}
\begin{multline}\label{solz}
    d_z(\textbf{k})\Delta_p(\textbf{k},\omega_k) = k_B T\times \\ \times\sum_{n,\textbf{k}'\neq0}( \tilde{g}_n(\textbf{k}',\omega_k-\omega_n)
    + \tilde{g}_s(\textbf{k}',\omega_k-\omega_n))\times\\ \times\Delta_p(\textbf{k}-\textbf{k}',\omega_n)\upsilon_z(\textbf{k}
    -\textbf{k}',\omega_n),
\end{multline}
\begin{multline}\label{sol0}
    \Delta_s(\textbf{k},\omega_k) = k_B T \times \\ \times\sum_{n,\textbf{k}'\neq0}\left(-\frac{g_F}{\Omega} + \tilde{g}_n(\textbf{k}',\omega_k
    -\omega_n) - \tilde{g}_s(\textbf{k}',\omega_k-\omega_n)\right)\times \\ \times\Delta_s(\textbf{k}
    -\textbf{k}',\omega_n)\varepsilon(\textbf{k}-\textbf{k}',\omega_n),
\end{multline}
{where}
\begin{multline}
    \bm{\upsilon}(\textbf{k},\Delta_p,\Delta_s)=\frac{1}{(\hbar^2\omega_n^2 + E_+^2(\textbf{k}))
    (\hbar^2\omega_n^2 + E_-^2(\textbf{k}))}\times \\ \times \Big\{ (\hbar^2\omega_n^2+\xi_{\textbf{k}}^2+\Delta^2(\textbf{k}))
    \textbf{d}(\textbf{k})-i\Delta_p^2(\textbf{k})[\textbf{m}(\textbf{k})\times \textbf{d}(\textbf{k})] +\\+
    2\Delta_s(\textbf{k})\Delta_p(\textbf{k}) \textbf{m}(\textbf{k}) +2\Delta_s^2(\textbf{k})\textrm{Re}[\textbf{d}(\textbf{k})]\Big\},
\end{multline}
\begin{multline}
    \varepsilon(\textbf{k},\Delta_p,\Delta_s)=\frac{1}{(\hbar^2\omega_n^2 + E_+^2(\textbf{k}))(\hbar^2\omega_n^2 + E_-^2(\textbf{k}))}\times\\
     \times \Big\{
    \hbar^2\omega_n^2+\xi_{\textbf{k}}^2+\Delta^2(\textbf{k})-2\Delta_p^2(\textbf{k})(\textrm{Re}[\textbf{d}(\textbf{k})]\cdot
    \textbf{d}(\textbf{k}))\Big\}
\end{multline}
\begin{gather*}
    \textbf{m} ( \textbf{k} )= i [\textbf{d}(\textbf{k})\times \textbf{d}^*(\textbf{k})],\\
    \Delta^2(\textbf{k}) = \Delta_p^2(\textbf{k})|d(\textbf{k})|^2+\Delta_s^2(\textbf{k}),\\
    E_{\pm}(\textbf{k}) =\sqrt{\xi_{\textbf{k}}^2 + \Delta^2 (\textbf{k})  \pm \Delta_p^2(\textbf{k})\left| \textbf{m}(\textbf{k})
    +2 \frac{\Delta_s(\textbf{k})}{\Delta_p(\textbf{k})}\textrm{Re}[\textbf{d(\textbf{k})}]\right| }.
\end{gather*}

The derived gap equations as it must be in the limiting case of vanishing bose-fermionic coupling ($\alpha\to 0$, $\beta\to 0$) yield the standard
BCS solution for attractive interactions in the fermionic sector ($g_F<0$). {For $\beta = \alpha = 0$ and $\textbf{d} = \textbf{0}$, the gap equations in the limit $T\rightarrow 0$ take the form}
\begin{multline*}
    \Delta = -\frac{ \Delta g_F}{2(2\pi)^3}\int\frac{d\textbf{k}}{\sqrt{\xi^2+\Delta^2}}= - \frac{ \Delta g_F M_F p_F}{2 \pi^2 \hbar^3}
    \ln \frac{\hbar\omega_F}{\Delta} ,
\end{multline*}
where $\hbar\omega_F$ is the  cut-off energy of fermions (the analougue of the Debye energy). {For $g_F > 0 $, the only possible solution is $\Delta = 0$. However,
if $g_F < 0$, we obtain the well known BCS-type result}
\begin{equation*}
    \Delta = \hbar\omega_F e^{-\frac{1}{\zeta}},
\end{equation*}
{where $\zeta = \frac{ | g_F | M_F p_0}{2 \pi^2 \hbar^3}.$}

 For a repulsive fermion-fermion interaction ($g_F\ge 0$),
the electron-electron pairing is possible only because of the effective interaction induced by Bogoluibov excitations in the bosonic
sector. As expected the density-density bose-fermi interaction (the term proportional to $\alpha^2$ in Eq. \eqref{sol0})
produces effective attractive interactions in the $s$-wave channel. At the same time the spin-spin bose-fermi interactions (the term
proportional to $\beta^2$ in Eq. \eqref{sol0}) lead to an additional \textit{repulsive} interaction for the $s$-wave channel.
Thus, in a sharp contrast with the $s$-wave superfluidity, for the $p$-wave superfluidity, both the spin-spin and
density-density bose-fermi interactions lead to an effective attraction that results in the realization of the $p$-wave
superfluid.

In the present work we consider solutions of the gap equations for the case $\Delta_s=0$ and $\Delta_p\ne 0$ since the $s$-wave channel can be eliminated  by tuning the interaction parameters both for $g_F<0$ and for $g_F>0$.  Furthermore,  for $g_F > 0$  it is easy to verify   that  while the $p$-wave pairing is readily  accessible, the $s$-wave pairing could be realized only if $\tilde{g}_n$ is sufficiently large. The gap equations for
the $s$-wave channel imply the following estimation of the interaction kernel:
\begin{multline}
    -\frac{g_F}{\Omega} + \tilde{g}_n(\textbf{k}',\omega_k-\omega_n) - \tilde{g}_s(\textbf{k}',\omega_k-\omega_n)\\\leq -\frac{g_F}{\Omega}
    + \tilde{g}_n(\textbf{k}',\omega_k-\omega_n)
    \leq -\frac{g_F}{\Omega}
    +\frac{\alpha^2}{4\Omega n_B c_0}.
\end{multline}
Therefore $s$-wave channel is essentially suppressed while $p$-wave channel is favoured in the systems with repulsive  ($g_F > 0$) fermion-fermion interactions.



\begin{thebibliography}{30}
\expandafter\ifx\csname natexlab\endcsname\relax\def\natexlab#1{#1}\fi
\expandafter\ifx\csname bibnamefont\endcsname\relax
  \def\bibnamefont#1{#1}\fi
\expandafter\ifx\csname bibfnamefont\endcsname\relax
  \def\bibfnamefont#1{#1}\fi
\expandafter\ifx\csname citenamefont\endcsname\relax
  \def\citenamefont#1{#1}\fi
\expandafter\ifx\csname url\endcsname\relax
  \def\url#1{\texttt{#1}}\fi
\expandafter\ifx\csname urlprefix\endcsname\relax\def\urlprefix{URL }\fi
\providecommand{\bibinfo}[2]{#2}
\providecommand{\eprint}[2][]{\url{#2}}

\bibitem[{\citenamefont{Nayak et~al.}(2008)\citenamefont{Nayak, Simon, Stern,
  Freedman, and Das~Sarma}}]{RevModPhys-Das-Sarma}
\bibinfo{author}{\bibfnamefont{C.}~\bibnamefont{Nayak}},
  \bibinfo{author}{\bibfnamefont{S.~H.} \bibnamefont{Simon}},
  \bibinfo{author}{\bibfnamefont{A.}~\bibnamefont{Stern}},
  \bibinfo{author}{\bibfnamefont{M.}~\bibnamefont{Freedman}}, \bibnamefont{and}
  \bibinfo{author}{\bibfnamefont{S.}~\bibnamefont{Das~Sarma}},
  \bibinfo{journal}{Rev. Mod. Phys.} \textbf{\bibinfo{volume}{80}},
  \bibinfo{pages}{1083} (\bibinfo{year}{2008}).

\bibitem[{\citenamefont{Ivanov}(2001)}]{ivanov2001non}
\bibinfo{author}{\bibfnamefont{D.~A.} \bibnamefont{Ivanov}},
  \bibinfo{journal}{Phys. Rev. Lett.} \textbf{\bibinfo{volume}{86}},
  \bibinfo{pages}{268} (\bibinfo{year}{2001}).

\bibitem[{\citenamefont{{Kitaev}}(2003)}]{Kitaev}
\bibinfo{author}{\bibfnamefont{A.~Y.} \bibnamefont{{Kitaev}}},
  \bibinfo{journal}{Annals of Physics} \textbf{\bibinfo{volume}{303}},
  \bibinfo{pages}{2} (\bibinfo{year}{2003}).

\bibitem[{\citenamefont{Moore and Read}(1991)}]{moore1991nonabelions}
\bibinfo{author}{\bibfnamefont{G.}~\bibnamefont{Moore}} \bibnamefont{and}
  \bibinfo{author}{\bibfnamefont{N.}~\bibnamefont{Read}},
  \bibinfo{journal}{Nuclear Physics B} \textbf{\bibinfo{volume}{360}},
  \bibinfo{pages}{362} (\bibinfo{year}{1991}).

\bibitem[{\citenamefont{Volovik}(2003)}]{volovik2003universe}
\bibinfo{author}{\bibfnamefont{G.~E.} \bibnamefont{Volovik}},
  \emph{\bibinfo{title}{The universe in a helium droplet}}, vol.
  \bibinfo{volume}{117} (\bibinfo{publisher}{Oxford University Press},
  \bibinfo{year}{2003}).

\bibitem[{\citenamefont{Kallin}(2012)}]{kallin2012chiral}
\bibinfo{author}{\bibfnamefont{C.}~\bibnamefont{Kallin}},
  \bibinfo{journal}{Reports on Progress in Physics}
  \textbf{\bibinfo{volume}{75}}, \bibinfo{pages}{042501}
  (\bibinfo{year}{2012}).

\bibitem[{\citenamefont{Charpentier et~al.}(2017)\citenamefont{Charpentier,
  Galletti, Kunakova, Arpaia, Song, Baghdadi, Wang, Kalaboukhov, Olsson, Tafuri
  et~al.}}]{charpentier2017induced}
\bibinfo{author}{\bibfnamefont{S.}~\bibnamefont{Charpentier}},
  \bibinfo{author}{\bibfnamefont{L.}~\bibnamefont{Galletti}},
  \bibinfo{author}{\bibfnamefont{G.}~\bibnamefont{Kunakova}},
  \bibinfo{author}{\bibfnamefont{R.}~\bibnamefont{Arpaia}},
  \bibinfo{author}{\bibfnamefont{Y.}~\bibnamefont{Song}},
  \bibinfo{author}{\bibfnamefont{R.}~\bibnamefont{Baghdadi}},
  \bibinfo{author}{\bibfnamefont{S.~M.} \bibnamefont{Wang}},
  \bibinfo{author}{\bibfnamefont{A.}~\bibnamefont{Kalaboukhov}},
  \bibinfo{author}{\bibfnamefont{E.}~\bibnamefont{Olsson}},
  \bibinfo{author}{\bibfnamefont{F.}~\bibnamefont{Tafuri}},
  \bibnamefont{et~al.}, \bibinfo{journal}{Nature communications}
  \textbf{\bibinfo{volume}{8}}, \bibinfo{pages}{2019} (\bibinfo{year}{2017}).

\bibitem[{\citenamefont{J.~Levinsen and Gurarie}(2008)}]{Cooper}
\bibinfo{author}{\bibfnamefont{N.~C.} \bibnamefont{J.~Levinsen}}
  \bibnamefont{and} \bibinfo{author}{\bibfnamefont{V.}~\bibnamefont{Gurarie}},
  \bibinfo{journal}{Phys. Rev. A} \textbf{\bibinfo{volume}{78}},
  \bibinfo{eid}{063616} (\bibinfo{year}{2008}).

\bibitem[{\citenamefont{Zhang et~al.}(2004)\citenamefont{Zhang, Van~Kempen,
  Bourdel, Khaykovich, Cubizolles, Chevy, Teichmann, Tarruell, Kokkelmans, and
  Salomon}}]{zhang2004p}
\bibinfo{author}{\bibfnamefont{J.}~\bibnamefont{Zhang}},
  \bibinfo{author}{\bibfnamefont{E.}~\bibnamefont{Van~Kempen}},
  \bibinfo{author}{\bibfnamefont{T.}~\bibnamefont{Bourdel}},
  \bibinfo{author}{\bibfnamefont{L.}~\bibnamefont{Khaykovich}},
  \bibinfo{author}{\bibfnamefont{J.}~\bibnamefont{Cubizolles}},
  \bibinfo{author}{\bibfnamefont{F.}~\bibnamefont{Chevy}},
  \bibinfo{author}{\bibfnamefont{M.}~\bibnamefont{Teichmann}},
  \bibinfo{author}{\bibfnamefont{L.}~\bibnamefont{Tarruell}},
  \bibinfo{author}{\bibfnamefont{S.}~\bibnamefont{Kokkelmans}},
  \bibnamefont{and} \bibinfo{author}{\bibfnamefont{C.}~\bibnamefont{Salomon}},
  \bibinfo{journal}{Phys. Rev. A} \textbf{\bibinfo{volume}{70}},
  \bibinfo{pages}{030702} (\bibinfo{year}{2004}).

\bibitem[{\citenamefont{Gaebler et~al.}(2007)\citenamefont{Gaebler, Stewart,
  Bohn, and Jin}}]{gaebler2007p}
\bibinfo{author}{\bibfnamefont{J.}~\bibnamefont{Gaebler}},
  \bibinfo{author}{\bibfnamefont{J.}~\bibnamefont{Stewart}},
  \bibinfo{author}{\bibfnamefont{J.}~\bibnamefont{Bohn}}, \bibnamefont{and}
  \bibinfo{author}{\bibfnamefont{D.}~\bibnamefont{Jin}},
  \bibinfo{journal}{Phys. Rev. Lett.} \textbf{\bibinfo{volume}{98}},
  \bibinfo{pages}{200403} (\bibinfo{year}{2007}).

\bibitem[{\citenamefont{Levinsen et~al.}(2008)\citenamefont{Levinsen, Cooper,
  and Gurarie}}]{levinsen2008stability}
\bibinfo{author}{\bibfnamefont{J.}~\bibnamefont{Levinsen}},
  \bibinfo{author}{\bibfnamefont{N.}~\bibnamefont{Cooper}}, \bibnamefont{and}
  \bibinfo{author}{\bibfnamefont{V.}~\bibnamefont{Gurarie}},
  \bibinfo{journal}{Phys. Rev. A} \textbf{\bibinfo{volume}{78}},
  \bibinfo{pages}{063616} (\bibinfo{year}{2008}).

\bibitem[{\citenamefont{Jona-Lasinio et~al.}(2008)\citenamefont{Jona-Lasinio,
  Pricoupenko, and Castin}}]{jona2008three}
\bibinfo{author}{\bibfnamefont{M.}~\bibnamefont{Jona-Lasinio}},
  \bibinfo{author}{\bibfnamefont{L.}~\bibnamefont{Pricoupenko}},
  \bibnamefont{and} \bibinfo{author}{\bibfnamefont{Y.}~\bibnamefont{Castin}},
  \bibinfo{journal}{Phys. Rev. A} \textbf{\bibinfo{volume}{77}},
  \bibinfo{pages}{043611} (\bibinfo{year}{2008}).

\bibitem[{\citenamefont{Cooper and Shlyapnikov}(2009)}]{cooper2009stable}
\bibinfo{author}{\bibfnamefont{N.}~\bibnamefont{Cooper}} \bibnamefont{and}
  \bibinfo{author}{\bibfnamefont{G.}~\bibnamefont{Shlyapnikov}},
  \bibinfo{journal}{Phys. Rev. Lett.} \textbf{\bibinfo{volume}{103}},
  \bibinfo{pages}{155302} (\bibinfo{year}{2009}).

\bibitem[{\citenamefont{Zhang et~al.}(2008)\citenamefont{Zhang, Tewari,
  Lutchyn, and Sarma}}]{zhang2008p}
\bibinfo{author}{\bibfnamefont{C.}~\bibnamefont{Zhang}},
  \bibinfo{author}{\bibfnamefont{S.}~\bibnamefont{Tewari}},
  \bibinfo{author}{\bibfnamefont{R.~M.} \bibnamefont{Lutchyn}},
  \bibnamefont{and} \bibinfo{author}{\bibfnamefont{S.~D.} \bibnamefont{Sarma}},
  \bibinfo{journal}{Phys. Rev. Lett.} \textbf{\bibinfo{volume}{101}},
  \bibinfo{pages}{160401} (\bibinfo{year}{2008}).

\bibitem[{\citenamefont{Sato et~al.}(2009)\citenamefont{Sato, Takahashi, and
  Fujimoto}}]{sato2009non}
\bibinfo{author}{\bibfnamefont{M.}~\bibnamefont{Sato}},
  \bibinfo{author}{\bibfnamefont{Y.}~\bibnamefont{Takahashi}},
  \bibnamefont{and} \bibinfo{author}{\bibfnamefont{S.}~\bibnamefont{Fujimoto}},
  \bibinfo{journal}{Phys. Rev. Lett.} \textbf{\bibinfo{volume}{103}},
  \bibinfo{pages}{020401} (\bibinfo{year}{2009}).

\bibitem[{\citenamefont{Williams et~al.}(2012)\citenamefont{Williams, LeBlanc,
  Jimenez-Garcia, Beeler, Perry, Phillips, and
  Spielman}}]{williams2012synthetic}
\bibinfo{author}{\bibfnamefont{R.~A.} \bibnamefont{Williams}},
  \bibinfo{author}{\bibfnamefont{L.~J.} \bibnamefont{LeBlanc}},
  \bibinfo{author}{\bibfnamefont{K.}~\bibnamefont{Jimenez-Garcia}},
  \bibinfo{author}{\bibfnamefont{M.~C.} \bibnamefont{Beeler}},
  \bibinfo{author}{\bibfnamefont{A.~R.} \bibnamefont{Perry}},
  \bibinfo{author}{\bibfnamefont{W.~D.} \bibnamefont{Phillips}},
  \bibnamefont{and} \bibinfo{author}{\bibfnamefont{I.~B.}
  \bibnamefont{Spielman}}, \bibinfo{journal}{Science}
  \textbf{\bibinfo{volume}{335}}, \bibinfo{pages}{314} (\bibinfo{year}{2012}).

\bibitem[{\citenamefont{Juli{\'a}-D{\'\i}az
  et~al.}(2013)\citenamefont{Juli{\'a}-D{\'\i}az, Gra{\ss}, Dutta, Chang, and
  Lewenstein}}]{julia2013engineering}
\bibinfo{author}{\bibfnamefont{B.}~\bibnamefont{Juli{\'a}-D{\'\i}az}},
  \bibinfo{author}{\bibfnamefont{T.}~\bibnamefont{Gra{\ss}}},
  \bibinfo{author}{\bibfnamefont{O.}~\bibnamefont{Dutta}},
  \bibinfo{author}{\bibfnamefont{D.}~\bibnamefont{Chang}}, \bibnamefont{and}
  \bibinfo{author}{\bibfnamefont{M.}~\bibnamefont{Lewenstein}},
  \bibinfo{journal}{Nature communications} \textbf{\bibinfo{volume}{4}},
  \bibinfo{pages}{2046} (\bibinfo{year}{2013}).

\bibitem[{\citenamefont{Han et~al.}(2009)\citenamefont{Han, Chan, Yi, Daley,
  Diehl, Zoller, and Duan}}]{han2009stabilization}
\bibinfo{author}{\bibfnamefont{Y.-J.} \bibnamefont{Han}},
  \bibinfo{author}{\bibfnamefont{Y.-H.} \bibnamefont{Chan}},
  \bibinfo{author}{\bibfnamefont{W.}~\bibnamefont{Yi}},
  \bibinfo{author}{\bibfnamefont{A.}~\bibnamefont{Daley}},
  \bibinfo{author}{\bibfnamefont{S.}~\bibnamefont{Diehl}},
  \bibinfo{author}{\bibfnamefont{P.}~\bibnamefont{Zoller}}, \bibnamefont{and}
  \bibinfo{author}{\bibfnamefont{L.-M.} \bibnamefont{Duan}},
  \bibinfo{journal}{Phys. Rev. Lett.} \textbf{\bibinfo{volume}{103}},
  \bibinfo{pages}{070404} (\bibinfo{year}{2009}).

\bibitem[{\citenamefont{Bardyn et~al.}(2012)\citenamefont{Bardyn, Baranov,
  Rico, {\.I}mamo{\u{g}}lu, Zoller, and Diehl}}]{bardyn2012majorana}
\bibinfo{author}{\bibfnamefont{C.-E.} \bibnamefont{Bardyn}},
  \bibinfo{author}{\bibfnamefont{M.}~\bibnamefont{Baranov}},
  \bibinfo{author}{\bibfnamefont{E.}~\bibnamefont{Rico}},
  \bibinfo{author}{\bibfnamefont{A.}~\bibnamefont{{\.I}mamo{\u{g}}lu}},
  \bibinfo{author}{\bibfnamefont{P.}~\bibnamefont{Zoller}}, \bibnamefont{and}
  \bibinfo{author}{\bibfnamefont{S.}~\bibnamefont{Diehl}},
  \bibinfo{journal}{Phys. Rev. Lett.} \textbf{\bibinfo{volume}{109}},
  \bibinfo{pages}{130402} (\bibinfo{year}{2012}).

\bibitem[{\citenamefont{Efremov and Viverit}(2002)}]{Efremov}
\bibinfo{author}{\bibfnamefont{D.~V.} \bibnamefont{Efremov}} \bibnamefont{and}
  \bibinfo{author}{\bibfnamefont{L.}~\bibnamefont{Viverit}},
  \bibinfo{journal}{Phys. Rev. B} \textbf{\bibinfo{volume}{65}},
  \bibinfo{pages}{134519} (\bibinfo{year}{2002}).

\bibitem[{\citenamefont{Wang et~al.}(2005)\citenamefont{Wang, Lukin, and
  Demler}}]{wang2005engineering}
\bibinfo{author}{\bibfnamefont{D.-W.} \bibnamefont{Wang}},
  \bibinfo{author}{\bibfnamefont{M.~D.} \bibnamefont{Lukin}}, \bibnamefont{and}
  \bibinfo{author}{\bibfnamefont{E.}~\bibnamefont{Demler}},
  \bibinfo{journal}{Phys. Rev. A} \textbf{\bibinfo{volume}{72}},
  \bibinfo{pages}{051604} (\bibinfo{year}{2005}).

\bibitem[{\citenamefont{Massignan et~al.}(2010)\citenamefont{Massignan,
  Sanpera, and Lewenstein}}]{massignan2010creating}
\bibinfo{author}{\bibfnamefont{P.}~\bibnamefont{Massignan}},
  \bibinfo{author}{\bibfnamefont{A.}~\bibnamefont{Sanpera}}, \bibnamefont{and}
  \bibinfo{author}{\bibfnamefont{M.}~\bibnamefont{Lewenstein}},
  \bibinfo{journal}{Phys. Rev. A} \textbf{\bibinfo{volume}{81}},
  \bibinfo{pages}{031607} (\bibinfo{year}{2010}).

\bibitem[{\citenamefont{Wu and Bruun}(2016)}]{wu2016topological}
\bibinfo{author}{\bibfnamefont{Z.}~\bibnamefont{Wu}} \bibnamefont{and}
  \bibinfo{author}{\bibfnamefont{G.~M.} \bibnamefont{Bruun}},
  \bibinfo{journal}{Phys. Rev. Letters} \textbf{\bibinfo{volume}{117}},
  \bibinfo{pages}{245302} (\bibinfo{year}{2016}).

\bibitem[{\citenamefont{Mathey et~al.}(2006)\citenamefont{Mathey, Tsai, and
  Neto}}]{mathey2006competing}
\bibinfo{author}{\bibfnamefont{L.}~\bibnamefont{Mathey}},
  \bibinfo{author}{\bibfnamefont{S.-W.} \bibnamefont{Tsai}}, \bibnamefont{and}
  \bibinfo{author}{\bibfnamefont{A.~C.} \bibnamefont{Neto}},
  \bibinfo{journal}{Phys. Rev. Lett.} \textbf{\bibinfo{volume}{97}},
  \bibinfo{pages}{030601} (\bibinfo{year}{2006}).

\bibitem[{\citenamefont{Ho}(1998)}]{PhysRevLett.81.742}
\bibinfo{author}{\bibfnamefont{T.-L.} \bibnamefont{Ho}},
  \bibinfo{journal}{Phys. Rev. Lett.} \textbf{\bibinfo{volume}{81}},
  \bibinfo{pages}{742} (\bibinfo{year}{1998}).

\bibitem[{\citenamefont{{Kawaguchi} and {Ueda}}(2012)}]{2012PhR...520..253K}
\bibinfo{author}{\bibfnamefont{Y.}~\bibnamefont{{Kawaguchi}}} \bibnamefont{and}
  \bibinfo{author}{\bibfnamefont{M.}~\bibnamefont{{Ueda}}},
  \bibinfo{journal}{Phys. Rep.} \textbf{\bibinfo{volume}{520}},
  \bibinfo{pages}{253} (\bibinfo{year}{2012}).

\bibitem[{\citenamefont{{Stamper-Kurn} and {Ueda}}(2013)}]{2013RvMP...85.1191S}
\bibinfo{author}{\bibfnamefont{D.~M.} \bibnamefont{{Stamper-Kurn}}}
  \bibnamefont{and} \bibinfo{author}{\bibfnamefont{M.}~\bibnamefont{{Ueda}}},
  \bibinfo{journal}{Rev. Mod. Phys.} \textbf{\bibinfo{volume}{85}},
  \bibinfo{pages}{1191} (\bibinfo{year}{2013}).

\bibitem[{\citenamefont{Pitaevskii and Stringari}(2016)}]{Pitaevskii2016}
\bibinfo{author}{\bibfnamefont{L.}~\bibnamefont{Pitaevskii}} \bibnamefont{and}
  \bibinfo{author}{\bibfnamefont{S.}~\bibnamefont{Stringari}},
  \emph{\bibinfo{title}{Bose-Einstein condensation and superfluidity}}
  (\bibinfo{publisher}{Oxford University Press}, \bibinfo{year}{2016}).

\bibitem[{\citenamefont{Mineev and Samokhin}(1999)}]{Min}
\bibinfo{author}{\bibfnamefont{V.}~\bibnamefont{Mineev}} \bibnamefont{and}
  \bibinfo{author}{\bibfnamefont{K.}~\bibnamefont{Samokhin}},
  \emph{\bibinfo{title}{Introduction to the theory of unusual
  superconductivity}} (\bibinfo{publisher}{CRC Press}, \bibinfo{year}{1999}).

\bibitem[{\citenamefont{Petviashvili and Yan'kov}(1989)}]{Petviashvili86}
\bibinfo{author}{\bibfnamefont{V.}~\bibnamefont{Petviashvili}}
  \bibnamefont{and} \bibinfo{author}{\bibfnamefont{V.}~\bibnamefont{Yan'kov}},
  \emph{\bibinfo{title}{Rev. Plasma Phys.}} (\bibinfo{publisher}{Consultants
  Bureau, New York}, \bibinfo{year}{1989}).

\end{thebibliography}

\end{document}